\documentclass[12pt]{iopart}

\usepackage{color}
\usepackage{graphicx}
\usepackage{xspace}
\usepackage[normalem]{ulem}
\usepackage{appendix}
\usepackage{hyperref}

\begin{document}
\newcommand{\levele}{$^3H_4$\xspace}
\newcommand{\levelh}{$^3H_5$\xspace}
\newcommand{\levelm}{$^3F_4$\xspace}
\newcommand{\levelg}{$^3H_6$\xspace}

\newcommand{\perroothz}{\,\mathrm{Hz}^{-1/2}}
\newcommand{\nact}{n_{\mathrm{active}}}
\newcommand{\Gstress}{G_{\mathrm{stress}}}
\newcommand{\Gammaeff}{\Gamma_{\mathrm{eff}}}

\newcommand{\Hzcm}{Hz\,cm$^3$}
\newcommand{\betaexp}{$\beta_{\mathrm{exp}}$\xspace}

\newcommand{\TC}[1]{{\color{magenta}{TC: #1}}}
\newcommand{\ALC}[1]{{\color{cyan}{ALC: #1}}}

\title{Strain-mediated ion-ion interaction in rare-earth-doped solids}

\author{A. Louchet-Chauvet}
\ead{anne.louchet-chauvet@espci.fr}
\address{ESPCI Paris, Université PSL, CNRS, Institut Langevin, 75005 Paris, France}
\author{T. Chanelière}
\address{Univ. Grenoble Alpes, CNRS, Grenoble INP, Institut Néel, 38000 Grenoble, France}

\begin{abstract}
It was recently shown that the optical excitation of rare-earth ions produces a local change of the host matrix shape, attributed to a change of the rare-earth ion's electronic orbital geometry. In this work we investigate the consequences of this piezo-orbital backaction and show from a macroscopic model how it yields a disregarded ion-ion interaction mediated by mechanical strain. This interaction scales as $1/r^3$, similarly to the other archetypal ion-ion interactions, namely electric and magnetic dipole-dipole interactions.
We quantitatively assess and compare the magnitude of these three interactions from the angle of the instantaneous spectral diffusion mechanism, and reexamine the scientific literature in a range of rare-earth doped systems in the light of this generally underestimated contribution.
\end{abstract}
\maketitle

\section{Introduction}
Atomic ensembles have attracted a lot of attention over more than 20 years due to their inherent capacity to efficiently interact with light~\cite{hammerer2010quantum}. They are at the center of a number of quantum storage protocols, in the form of gases~\cite{dlcz2001,lukin2003} or solid state~\cite{tittel2010photon,grezes2016towards}. Among the most interesting solid state candidates, rare-earth ion-doped crystals (REIC) are particularly attractive due to their long optical coherence lifetimes at cryogenic temperatures~\cite{thiel2011rare} and are at the center of a number of actively developed quantum memory protocols~\cite{liu2021heralded,duda2022optimising,minnegaliev2022implementation}. In these light-matter interfaces, the optical depth of the medium is an important figure of merit since it enables high storage and retrieval efficiency~\cite{gorshkov2006universal,afzelius2009multimode,grezes2016towards}. However, reaching large optical depths usually comes with working with large atomic concentrations, leading to reinforced ion-ion interactions and thereby enhanced decoherence~\cite{agnello2001instantaneous,bottger2006,dajczgewand2015,lim2018coherent,alexander2022highcoop,rancic2022electron}.

Interestingly, although ion-ion interactions are potentially detrimental to the performance of quantum devices, they have also emerged as the foundational mechanism for quantum computing architectures because they allow multiqubit gate operations~\cite{ladd2010quantum}. Quantum computing was initially developed in physical systems with strong readout capacity ranging from dilute systems (trapped atoms) to condensed matter (quantum dots or superconducting qubits). Rare-earth ion-doped crystals were only recently considered as relevant for such applications~\cite{kinos2021roadmap} thanks to the elaboration of efficient single ion readout schemes that compensate for the optical transition's weak oscillator strength~\cite{zhong2018optically,raha2020optical}.

Either way, proper understanding and quantifying of ion-ion interactions in REIC are crucial. In most systems, the order of magnitude of the measured interaction strength is compatible with magnetic and/or electric dipole-dipole interaction. In both mechanisms, the excitation of some ions induces a change in their electric or magnetic dipole moment, modifying the local field accordingly in their vicinity. This modified field affects the surrounding ions in proportion to the Stark or Zeeman sensitivity of their energy levels.
In a limited number of REIC, however, the actual magnitude of this interaction is larger than expected by several orders of magnitude~\cite{ahlefeldt2013precision,attalPhD}. 
In this paper we consider a strain-mediated ion-ion interaction that has not been investigated so far and that may explain this discrepancy. The interaction we consider stems from the apparition of an excitation-induced stress field, affecting the surrounding ions via their piezospectroscopic sensivity. This fundamental effect has been pointed out early in the context of paramagnetic resonance under RF excitation, named virtual phonon exchange interaction initially considering transition metals ions~\cite{sugihara1959spin, mcmahon1964virtual, Orbach_PhysRev.158.524}. At the time, the description exploited the equivalent operators formalism for the spin-lattice coupled system, but the different modeling parameters are difficult to infer from experimental measurements. Despite a series of studies including rare-earth salts \cite[and references]{CONE1987481}, phonon-mediated interactions seem to have been overlooked for years, before reappearing in the context of quantum technologies with original proposals of phononic engineering \cite{PhysRevLett.110.156402,PhysRevLett.120.213603}.

This paper is organized as follows. In Sec.~\ref{sec:strainmed}, the physical origin of this interaction is presented. A scaling law is given and an estimation of its strength is provided. In Sec.~\ref{sec:Comp3}, we quantitatively compare the magnitude of this strain-mediated interaction with respect to the electromagnetic dipole-dipole interactions in a number of host-dopant combinations and confront our predictions with the experimental data found in the scientific literature. The strength of the instantaneous spectral diffusion mechanism is used as a comparison criterion. Finally,  in Sec.~\ref{sec:discussion}, we discuss the implications of this work in quantum technology-related applications.

\section{Strain-mediated ion-ion interaction}
\label{sec:strainmed}
\subsection{General description}

When an atomic particle is promoted to a different electronic level, its outer shape and size are bound to change due to the modification of its electronic wavefunction. If the particle is embedded in a solid matrix, this \emph{piezo-orbital backaction} effect will additionally give rise to a stress field around the excited particle, making it detectable via a change of shape of the solid itself. This was recently evidenced in a bulk rare-earth ion-doped crystal in which only a finite volume of the crystal was illuminated, leading to a distortion of the nearby crystal surface~\cite{louchet2021optomechanical}.

Conversely, it has been known for several decades that the optical lines of rare-earth ions in solids are sensitive to stress via the piezospectroscopic effect~\cite{kaplyanskii1962piezospectroscopic,bungenstock2000effect}. Indeed, in the elastic regime, a compressive or tensile stress modifies the interatomic distances. This affects the crystal field and in turn shifts the rare-earth ion's energy levels. This sensitivity to stress was recently proposed as a tool to sense mechanical vibrations in a cryogenic environment~\cite{louchet2019piezospectroscopic,louchet2022limits}.

The \emph{piezo-orbital backaction} and the piezospectroscopic effect are in fact two facets of the same coupling mechanism. Their combination leads to a shift of the transition frequency in the ions surrounding a given excited rare-earth ion. In the following we coin this as the \emph{strain-mediated ion-ion interaction}.



\subsection{Magnitude of the strain-mediated interaction}

\begin{figure}[t]
\centering
\includegraphics[width=7.5cm]{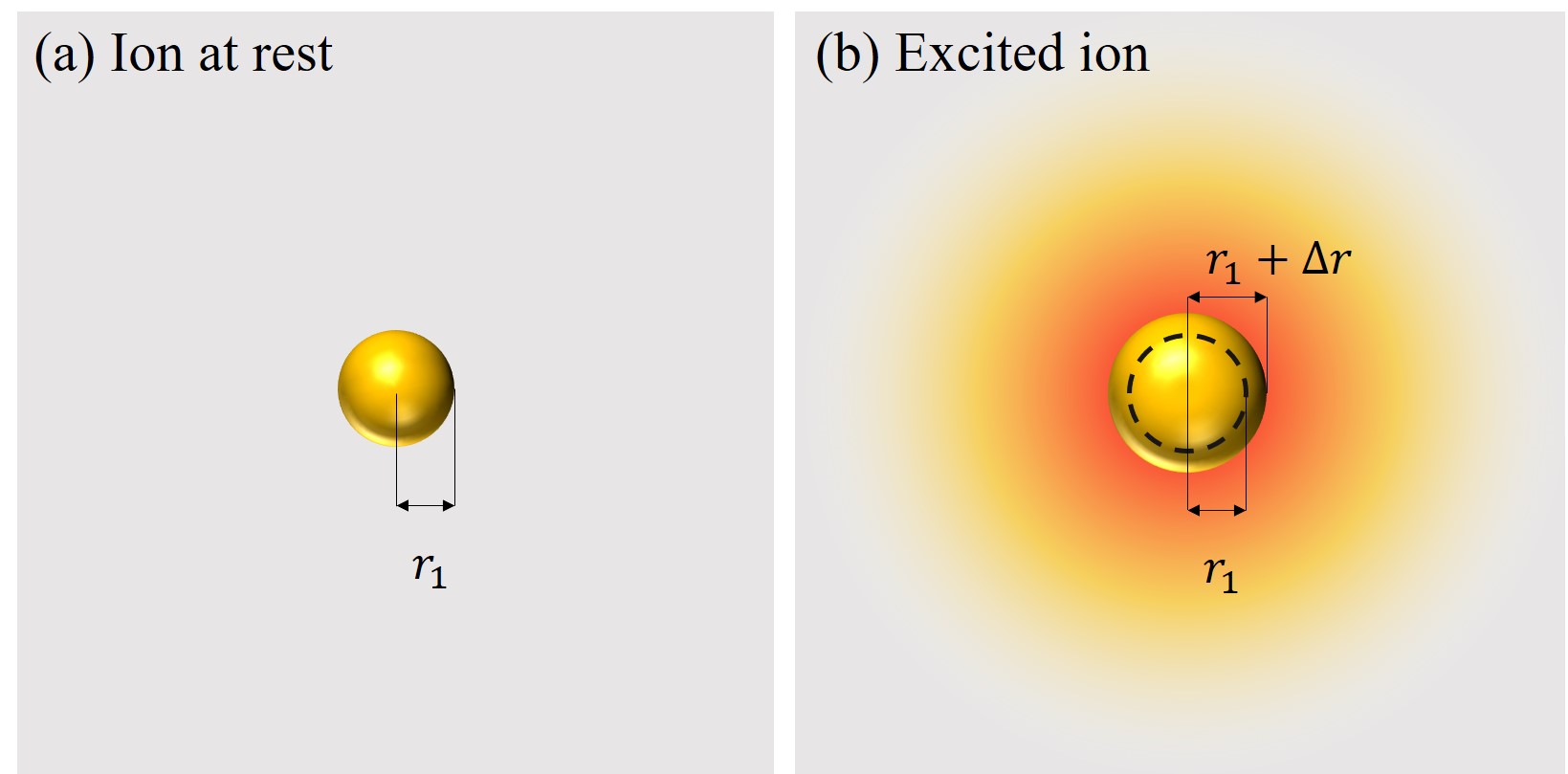}
\caption{Simplified view of the piezo-orbital backaction around a spherical rare-earth ion. The stress field is symbolized by a radial color gradient around the ion.}
\label{fig:StressField}
\end{figure}

For simplicity we assume that the ion is spherical with a radius $r_1$ and that the piezo-orbital backaction acts as a simple ionic radius change (see Fig.~\ref{fig:StressField}). This impacts the surrounding matrix by creating a radial stress field $\sigma(r)$ around the ion. With a spherically symmetric continuum mechanics model detailed in \ref{app:mechanics}, we calculate the elastic strain energy that is necessary to establish this stress field. Due to energy conservation, this elastic energy corresponds to the energy shift of the electronic levels due to the internal stress within the ionic volume (see \ref{app:energy}). This allows us to relate the ionic radius variation $\Delta r$ to the piezospectroscopic sensitivity $\kappa$:
\begin{equation}
\Delta r= \frac{h \kappa}{4 \pi r_1^2}
\label{eq:Deltar}
\end{equation}
where $h$ is the Planck constant. With this we derive the radial stress field $\sigma(r)$ and write the strain-mediated ion-ion interaction energy between two ions separated by a distance $r$:
\begin{equation}
E_{\mathrm{str}}(r)= h \kappa \sigma(r) = \frac{E}{1+\nu} \frac{(h\kappa)^2}{2\pi r^3}
\end{equation}
where $E$ is the Young modulus and $\nu$ is the Poisson's ratio of the crystal. We emphasize the relatively strong hypotheses underlying this result: the piezo-orbital backaction is assumed to occur in the form of a mere ionic radius change of the spherical excited ion, and the piezospectroscopic sensitivity is assumed to be scalar. The anisotropic nature of the crystalline matrix may naturally translate into a non-spherical strain field. Because of the large distance between dopants (larger than the crystal cell parameter at low concentration levels), we may expect the strain anisotropy to be weak. On the contrary, the piezospectroscopic sensitivity is usually described by a tensor~\cite{grabner1978spectroscopic}, so its scalar nature appears as a crude assumption in order to derive an order of magnitude.

Using Eq.~\ref{eq:Deltar} we can estimate the corresponding relative ionic radius change due to the piezo-orbital backaction in rare-earth-doped crystals. Taking typical values $\kappa=100$~Hz/Pa~\cite{louchet2019piezospectroscopic} and $r_1=1$~\AA (corresponding to the typical effective ionic radii for matrix-embedded rare-earth ions~\cite{shannon1976revised}), we obtain $\Delta r/r_1\simeq 5\cdot 10^{-3}$. This value remarkably agrees with the relative ionic radius change of a free ion among its $4f$ states that can be derived from a Hartree-Fock calculation (see~\cite{freeman1962theoretical} for Ce$^{3+}$ and~\cite{wybourne2007} for Eu$^{3+}$) even if the exact rearrangement of the outer shell defining the ionic radius surrounding the $4f$-shell deserves further analysis.

\section{Comparing the three ion-ion interactions}
\label{sec:Comp3}

In this section we propose to study how this interaction compares with the other usual ion-ion interactions generally considered in rare-earth doped systems, \emph{i.e.} electric and magnetic dipole-dipole interactions. We also confront these estimations to published measurements of ion-ion interactions in rare-earth doped crystals.

Interestingly, the strain-mediated interaction scales as $1/r^3$, exactly like the electric and magnetic dipole-dipole interactions, although originating from a radically different physical mechanism. All three interactions can be characterized by a coefficient $A_i$ such that their energies read as:
\begin{equation}
E_i(r)=\frac{h A_i}{2\pi r^3}
\end{equation}
where the $\{A_i\}$ are defined by the following:
\begin{eqnarray}
A_{\mathrm{el}}  &=& 2h\frac{1}{\epsilon_0 \epsilon_r} \Delta\mu_\mathrm{el}^2\label{eq:Ael}\\
A_{\mathrm{mag}} &=& 2h \mu_0  \Delta \mu_\mathrm{mag}^2 \label{eq:Amag}\\
A_{\mathrm{str}} &=& \frac{E}{1+\nu} h\kappa^2\label{eq:Astr}
\end{eqnarray}
$\Delta \mu_\mathrm{el}$ and $\Delta \mu_\mathrm{mag}$ are the electric and magnetic dipole moment variation when the ion is promoted to the excited state. They are expressed in Hz\,m\,V$^{-1}$ and in Hz\,T$^{-1}$, respectively.
We note that in all three expressions, the sensitivity of the optical transition to electric field, magnetic field or strain (respectively) appears as a square law, in agreement with what is expected in an ion-ion interaction.

This scaling has been early derived for paramagnetic impurities under RF excitation~\cite{sugihara1959spin,mcmahon1964virtual}, in the so-called zero-retardation limit of the virtual phonon exchange (VPE) interaction~\cite{Orbach_PhysRev.158.524}. At the time, McMahon {\it et al.} also showed that VPE and magnetic interactions have the same order of magnitude for transition metal paramagnetic dopants  \cite{mcmahon1964virtual}.

With these expressions one can anticipate the variability  of the three considered interactions amongst REIC. For example, the strain-mediated interaction should be quite constant across different ions or crystals, since the mechanical properties of typical crystalline rare-earth doped oxides and the piezospectroscopic sensitivity of their optical lines are rather similar~\cite{louchet2019piezospectroscopic}.
On the other hand, the magnetic dipole-dipole interactions vary by several orders of magnitude between Kramers and non-Kramers ions because some exhibit an electronic spin while some only have a nuclear spin behaviour. A variability of 6 orders of magnitude is expected, since $(\mu_B/\mu_N)^2\simeq 3\cdot10^6$, where $\mu_B$ and $\mu_N$ are the Bohr magneton and the nuclear magneton, respectively. A large variability is also expected for the electric dipole-dipole interaction because the appearance of an electric dipole is only permitted by the crystal field that weakly perturbs the free ion electronic structure. Indeed different hosts, depending on the crystal site symmetry, allow or inhibit permanent electric dipole moment for the rare earth ion dopant.

The overall ion-ion interactions taking place in an ensemble of rare-earth ions can be probed experimentally by the observation of instantaneous spectral diffusion (ISD)~\cite{salikhov1981theory}, \emph{ie} the effect of a substantial population transfer on the spectral width of a narrow subset of ions within a given volume.
Since the ions are randomly positioned within the host matrix, the local modification of their environment following the excitation of one of them is not uniform. This results in a combination of an overall shift and dispersion of their resonance frequencies, among which only the latter leads to decoherence that is routinely measured. This is why the effect of ISD is described by an additional term to the homogeneous linewidth $\Gamma_h$. When it builds upon interactions scaling as $1/r^3$, this term
is proportional to the volumic density of excited particles $n_e$:
 \begin{equation}
\Gammaeff=\Gamma_h + \frac{\beta}{2}n_e
\end{equation}
where $\beta=\sum_i \beta_i$ is the sum of individual contributions due to each considered interaction~\cite{mims1972electron,dajczgewand2015}:
\begin{equation}
\label{eq:betaA}
\beta_i=\frac{8 \pi}{9\sqrt{3}} 10^6 A_i,
\end{equation}
the $10^6$ factor stemming from the choice of units (m$^3$\,s$^{-1}$ for $A_i$ and Hz\,cm$^3$ for $\beta_i$).


ISD is generally observed via high-resolution spectroscopy experiments such as photon echoes~\cite{taylor1974photon,liu1990,thiel2014isd}.
It has been evidenced in a large number of rare-earth ion-doped materials, but a quantitative value is only given in a handful of them. The published experimental values for $\beta$ in ten different rare-earth ion-doped crystals are given in Table~\ref{tab:betaexp}. These materials represent a rather complete sampling of the REIC diversity, including non-Kramers and Kramers ions, different crystal and site symmetries, and crystals with and without charge compensation. The measured values of $\beta$ are contained within a 2 order-of-magnitude range (between $10^{-13}$ and $10^{-11}$ Hz$\ $cm$^3$).

\begin{table}[ht]
\centering
\begin{tabular}{|c c|}
  \hline
  Cristal  &  \betaexp (\Hzcm)  \\
   \hline
  Tm:YAG               & $2.3\cdot 10^{-12}$~\cite{thiel2014isd}\\
  Tm:YGG               & $2.6\cdot 10^{-13}$~\cite{thiel2014isd}\\
  Tm:LiNbO$_3$         & $1.0\cdot 10^{-11}$~\cite{thiel2014isd}\\
  Er:YSO               & $1.3\cdot 10^{-12}$~\cite{thiel2012optical}\\
  Er:LiNbO$_3$         & $1.0\cdot 10^{-13}$~\cite{SMQIOA} \\
  Eu:YSO               & $9.0\cdot 10^{-13}$~\cite{konz2003} \\
  Eu:Y$_2$O$_3$        & $1.3\cdot 10^{-13}$~\cite{thiel2015qlims} \\ 
  EuCl$_3\cdot$6D$_2$O & $4.6\cdot 10^{-13}$~\cite{ahlefeldt2013precision} \\
  Pr:YSO               & $1.2\cdot 10^{-11}$~\cite{zhang2010novel}\\
  Pr:La$_2$(WO$_4$)$_3$ & $7.0\cdot 10^{-12}$~\cite{guillotnoel2007pr}\\
  \hline
\end{tabular}
\caption{Measured values for the ISD coefficient \betaexp in a selection of rare-earth ion-doped crystals.}
\label{tab:betaexp}
\end{table}

\begin{figure*}[ht]
\centering
\includegraphics[width=16cm]{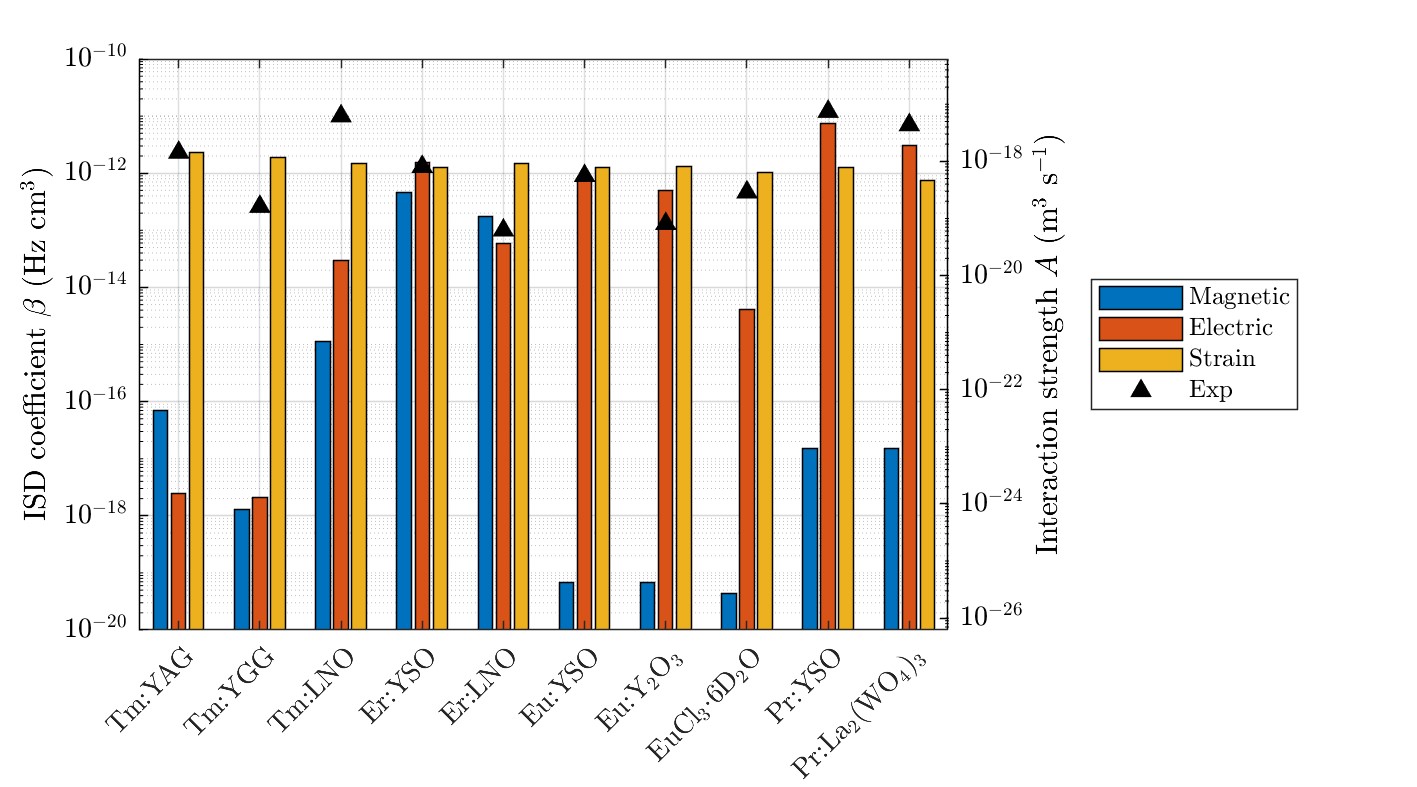}
\caption{Calculated (bars) and measured (triangles) values for $\beta_i$  and $A_i$ (where $i=\{\mathrm{mag}, \mathrm{el},\mathrm{str}, \mathrm{exp}\}$) for the 10 rare-earth ion-doped crystals for which quantitative values for \betaexp are available (see Table~\ref{tab:betaexp}). Note the logarithmic vertical scale spanning 10 orders of magnitude. }
\label{fig:ISD_All}
\end{figure*}

The three ion-ion interactions considered in this work (see equations~\ref{eq:Ael}, \ref{eq:Amag} and \ref{eq:Astr}) should contribute to ISD. Focusing our attention to the selection of REICs for which quantitative ISD measurements are available, we calculate the $\beta_i$ parameters for each ion-ion interactions using Eq.~\ref{eq:betaA} and display the results graphically in Figure~\ref{fig:ISD_All}.
The material parameters relevant to this calculation and resulting numerical estimations are given in \ref{sec:param}. Again, we point out that the values are mostly indicative since they rely on a very simplified modelling of the interactions.

We verify that both magnetic and electric dipole-dipole interactions vary among the materials within a 6 order-of-magnitude span depending on the existence of an electronic spin and the presence of a centrosymmetry in the doping site ($\beta_\mathrm{mag}$ is comprised between $4\cdot 10^{-20}$ and $5\cdot10^{-13}$~\Hzcm and $\beta_\mathrm{el}$ between $2\cdot10^{-18}$ and $8\cdot 10^{-12}$~\Hzcm, respectively). Conversely, the strain-mediated interaction is comprised within a much smaller interval ($\beta_\mathrm{str}$ between $7\cdot10^{-13}$ and $2\cdot10^{-12}$~\Hzcm) since we use a typical value for the piezospectroscopic sensitivity $\kappa$ for all and because of the very similar values of the Young moduli between crystals.

It is interesting to note that the strain-mediated interaction is of the order of the largest of the electric or magnetic dipole-dipole contributions found among all materials.
This means that when one or both of the dipole-dipole interactions are strong, they should coexist with the strain-mediated interaction. On the other hand, when both electromagnetic contributions are in the low range (weak Stark effect and no electronic spin for the non-Kramers ions, e.g. Tm-doped garnets or in a lesser part stoechiometric europium chloride), the strain-mediated interaction dominates by several orders of magnitude. In all considered REICs, the sum of the three predicted effects satisfactorily accounts for the measured values of \betaexp, finally providing an explanation to the previously large discrepancy between theoretical estimations and measurements of the ISD mechanism in some media.


\section{Discussion}
\label{sec:discussion}
Besides providing a better understanding of the physical origin and strength of ISD in REIC, the strain-mediated interaction in REIC could have interesting applications in the field of quantum computing. Indeed, quantum computing schemes rest upon the existence of a long-range atom-atom interaction enabling multiqubit operations~\cite{ladd2010quantum}. This interaction is often spontaneously assimilated to dipole-dipole interaction, and particularly in REIC~\cite{wesenberg2007scalable,kinos2021roadmap}.
We argue that the strain-mediated interaction could play this role in quantum-computing to replace the dipole-dipole coupling.
Such a scheme could be performed in almost any rare-earth ion-doped crystal since the strength of the interaction is rather similar over a broad variety of ion and host combinations~\cite{louchet2019piezospectroscopic}. In particular, the possession of a permanent electric dipole moment would not be an exclusive criteria for quantum computing compatibility. On the contrary, in some crystals (namely Tm-doped garnets, although there could be other candidates) this strain-mediated interaction dominates its electromagnetic counterparts by at least 4 orders of magnitude. This means that not only is the ion-ion interaction particularly pure in such media, but these strain-coupled qubits would be more robust against other types of decoherence process, such as magnetic field fluctuations due to spin flips in the host matrix~\cite{fraval2004, bottger2006}, or electric field noise due to charge fluctuations that may occur at the surface of rare-earth-doped nanoparticles~\cite{bartholomew2017optical}.

One may argue that since the strain-mediated interaction is intrinsically slow since it relies on the propagation of stress at the speed of sound (between $3000$ and $8000$~m/s in most considered crystals). We estimate its propagation delay by considering the average ion-ion distance $d_\mathrm{ion-ion}$, given by $\sqrt[3]{n_{\mathrm{RE}}}$ (where $n_{\mathrm{RE}}$ is the volumic density of rare-earth ions). $d_\mathrm{ion-ion}$  ranges from a few nm in highly concentrated materials (e.g. $\sim2$~nm in $1\%$ doped YAG) up to hundreds of nm in low concentration materials (e.g. $\sim650$~nm in $200$ppm doped YSO). This leads to a strain propagation time shorter than the nanosecond, to be compared with typical $\mu$s-scale light-matter interactions occuring in such media. Therefore, the strain-mediated interaction can still be considered instantaneous within rare-earth-doped crystals with typical concentrations.

\section{Conclusion}
Based on recent evidence of a conservative optomechanical backaction mechanism in rare-earth ion-doped crystals, we have unveiled a disregarded ion-ion interaction based on a physical mechanism fundamentally different from generally considered electromagnetic dipole-dipole interactions: the sensitivity of rare-earth ions to piezo-orbitally induced stress. With a simple mechanical model, we have estimated the strength of this strain-mediated interaction and shown that it is largely dominant in some rare-earth ion-doped crystals, opening interesting perspectives for strain-based quantum computing in solids.

\section{Acknowledgments}
The authors are grateful to Lars Rippe and Xiaoping Jia for helpful discussions. The authors acknowledge support from the French National Research Agency (ANR) through the projects ATRAP (ANR-19-CE24-0008), MIRESPIN (ANR-19-CE47-0011) and MARS (ANR-20-CE92-0041). This work has received support under the program  ``Investissements d’Avenir'' launched by the French Government.

\appendix

\section{Continuum mechanics in a REIC}
\label{app:mechanics}

\subsection{Spherical defect in an isotropic elastic medium}
Let us consider a sphere with radius $r_1$, embedded in an infinite, isotropic, continuous elastic medium with a Young modulus $E$ and a Poisson's ratio $\nu$. When a homogeneous radial stress $\sigma_0$ is applied in the sphere, the displacement field at a distance $r$ outside the sphere is radial and obeys~\cite{landau1967elast}:
\begin{equation}
u(r)=\sigma_0 \frac{1+\nu}{2E} \frac{r_1^3}{r^2},
\end{equation}
while the stress field around the sphere is also radial and reads as:
\begin{equation}
\sigma(r)=\sigma_0 \frac{r_1^3 }{r^3} \textrm{ for } r>r_1
\label{eq:sigma_cav}
\end{equation}

Defining $\Delta r = u(r_1)$ as the radius change, we obtain a simple relationship between $\sigma_0$ and $\Delta r$:
\begin{equation}
\sigma_0=\frac{\Delta r}{r_1} \frac{2E}{1+\nu}
\label{eq:Deltar_sigma0}
\end{equation}

\begin{figure}[ht]
\centering
\includegraphics[width=7.5cm]{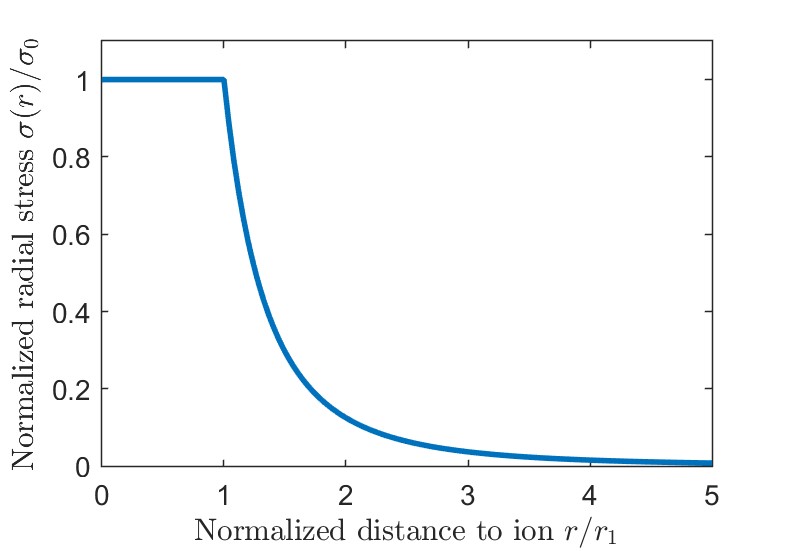}
\caption{Radial stress $\sigma(r)$ corresponding to the dilation of a sphere within an infinite isotropic medium. The stress is homogeneous within the sphere and decays with a $1/r^3$ law outside.}
\label{fig:Sigma}
\end{figure}

The radial stress can then be written:
\begin{equation}
 \sigma(r)=-\frac{\Delta r}{r_1} \frac{2E}{1+\nu} \frac{r_1^3 }{r^3}
 \label{eq:sigmarr}
\end{equation}

\subsection{Strain energy}
We now want to assess the elastic strain energy contained in the medium when the stress $\sigma_0$ is applied. We successively calculate the energy contained inside and outside the sphere~\cite{GereMechanics}.
\paragraph{Inside the sphere:}
The strain energy inside the sphere reads as:
\begin{equation}
U_\mathrm{int}=\frac 1 2 \sigma_0 \varepsilon_0 V
\end{equation}
where $\varepsilon_0=\Delta V/V=3\Delta r/r_1$ is the volumic change of the sphere. We finally obtain:
\begin{equation}
U_\mathrm{int}= 2\pi \Delta r \sigma_0  r_1^2
\label{eq:Uint}
\end{equation}

\paragraph{Outside the sphere:} The sphere being included in an infinite medium, we must also consider the strain energy that was necessary to establish the whole stress field around the sphere.
The volumic energy outside the sphere at a distance $r$ reads as:
\begin{equation}
u_V(r)=\frac 1 2 \sigma(r) \varepsilon(r)
\end{equation}
The volumic strain $\varepsilon(r)$ is related to the local stress via the volumic Hooke's law $\sigma(r)=K\varepsilon(r)$ (where $K=\frac{E}{1-2\nu}$ is the bulk modulus). Using Eq.~\ref{eq:sigma_cav} we get $\varepsilon(r)=\varepsilon_0r_ 1^3/r^3$, which finally leads to:
\begin{equation}
u_V(r)=\frac {3 \Delta r}{2} \sigma_0 \frac{r_1^5}{r^6}
\end{equation}
We integrate this volumic energy over the infinite volume of the medium:
\begin{equation}
U_\mathrm{ext} =4\pi \int_{r=r_1}^\infty u_V(r) r^2 dr = 6\pi \Delta r \sigma_0 r_1^5   \int_{r=r_1}^\infty \frac{1}{r^4} dr
\end{equation}
and obtain:
\begin{equation}
U_\mathrm{ext} =2\pi  \Delta r \sigma_0 r_1^2
\label{eq:Uext}
\end{equation}

\paragraph{Total strain energy:} Based on Eqs.~\ref{eq:Uint} and \ref{eq:Uext}, we derive the total strain energy that is necessary to distort the sphere and generate the associated stress field around it:
\begin{equation}
U_{\mathrm{strain}}=U_\mathrm{int}+U_\mathrm{ext}= 4\pi \Delta r \sigma_0  r_1^2
\label{eq:Utotale}
\end{equation}

\section{Relating the strain energy with the atomic energy}
\label{app:energy}
Let us now focus on the mechanical consequences of a change of state in a matrix-embedded rare-earth ion. Under optical excitation, due to piezo-orbital backaction~\cite{louchet2021optomechanical}, the rearrangement of the electronic orbitals is expected to alter the ion's apparent size.

In the following, we model the ion as an elastic sphere embedded in an elastic medium, and assume the piezo-orbitally-induced change of shape is merely a change of radius of this sphere. This way we may apply the calculation presented in \ref{app:mechanics}.
Due to energy conservation, this elastic energy is taken from the ion's energy levels. We can therefore write the following, considering that the shift in the ion's energy levels $\Delta E$ is related to the stress within the ion $\sigma_0$.
\begin{equation}
U_\mathrm{strain}=\Delta E = h \kappa \sigma_0
\label{eq:Udecalage}
\end{equation}
where $\kappa$ is the piezospectroscopic sensitivity, assumed scalar.
Injecting the expression of $U_\mathrm{strain}$ given by Eq.~\ref{eq:Utotale}, we get:
\begin{equation}
\Delta r= \frac{h \kappa}{4 \pi r_1^2}
\label{eq:Deltar_app}
\end{equation}
This allows us to obtain a simple relationship between the radial stress $\sigma(r)$ and the ionic radius change $\Delta r$, using Eq.~\ref{eq:sigmarr}:
\begin{equation}
\sigma(r)= \frac{2E}{1+\nu} \frac{h\kappa}{4\pi r^3}
\label{eq:sigmaDeltar}
\end{equation}

\begin{table*}[t]
\centering
\begin{tabular}{|cccccc|}
  \hline
  Crystal  & $E$ & $\nu$ &  $\Delta \mu_{mag}$ & $\epsilon_r$ &  $\Delta\mu_{\mathrm{el}}$ \\
  \hline
  Tm:YAG & $270$~GPa~\cite{huang2012elastic} & $0.256$~\cite{huang2012elastic} & $320$~MHz/T~\cite{louchet2007} & $10.6$~\cite{hofmeister1992infrared}& $65$~Hz\ cm/V~\cite{minnegaliev2021linear}\\
  Tm:YGG & $224$~GPa~\cite{monteseguro2013electronic} & $0.28$~\cite{monteseguro2013electronic} &  $44$~MHz/T~\cite{thiel2014ygg} & $12$~\cite{hofmeister1992infrared} & $\star65$~Hz\ cm/V \\
  Tm:LiNbO$_3$ & $170$~GPa~\cite{Dielectric_LNO} & $0.25$~\cite{Dielectric_LNO}  & $1.3$~GHz/T~\cite{thiel2010tmlno} & $\dagger65$~\cite{Young_LNO} & $18$~kHz\ cm/V~\cite{thiel2010tmlno}\\
  Er:YSO & $150$~GPa~\cite{mirzai2021first} & $0.26$~\cite{mirzai2021first}  & $26$~GHz/T~\cite{sun2008magnetic} & $10$~\cite{carvalho2015multi} & $50$~kHz\ cm/V~\cite{rancicPhD} \\
  Er:LiNbO$_3$ & $170$~GPa~\cite{Dielectric_LNO} & $0.25$~\cite{Dielectric_LNO}  & $16$~GHz/T~\cite{thiel2010erlno} & $\dagger65$~\cite{Young_LNO} & $25$~kHz\ cm/V~\cite{hastings2006controlled}\\
  Eu:YSO & $150$~GPa~\cite{mirzai2021first} & $0.26$~\cite{mirzai2021first}  & $10$~MHz/T~\cite{arcangeli2014spectroscopy} & $10$~\cite{carvalho2015multi} & $35$~kHz\ cm/V \cite{graf1998investigations,macfarlane2014optical} \\
  Eu:Y$_2$O$_3$ & $\star120$~GPa & $\star0.25$  & $\star10$~MHz/T & $15$~\cite{robertson2004high} & $\star35$~kHz\ cm/V\\
  EuCl$_3\cdot$6D$_2$O & $\star120$~GPa & $\star0.25$  & $8$~MHz/T~\cite{ahlefeldt2013PhD} & $3.6$~\cite{ahlefeldt2013precision} & $1.57$~kHz\ cm/V~\cite{ahlefeldt2013precision}  \\
  Pr:YSO & $150$~GPa~\cite{mirzai2021first} & $0.26$~\cite{mirzai2021first}  & $150$~MHz/T~\cite{heinze2011control} & $10$~\cite{carvalho2015multi} & $111$~kHz\ cm/V \cite{nilsson2005} \\
  Pr:La$_2$(WO$_4$)$_3$  & $90$~GPa~\cite{najafvandzadeh2020first} & $0.3$~\cite{najafvandzadeh2020first} & $\star150$~MHz/T & $20$~\cite{young1973compilation} & $\star100$~kHz\ cm/V\\
\hline
\end{tabular}
\caption{Material parameters used to compute the interaction strengths. $E$ is the mechanical Young modulus, $\nu$ the Poisson's ratio, and $\epsilon_r$ is the dielectric constant of the host matrix. $\Delta\mu_{\mathrm{mag}}$ and $\Delta\mu_{\mathrm{el}}$ are the difference in the rare earth magnetic or electric moments between ground and excited states. When no values could be found in the literature, we chose a value among similar materials and indicated this with the symbol "$\star$". $\dagger$: The lithium niobate (LNO) crystal is anisotropic and exhibits different values of $\epsilon_r$ depending on the crystallographic direction~\cite{Young_LNO}. For simplicity we choose an intermediate value. }
\label{tab:param}
\end{table*}

\begin{table}[t]
\centering
\begin{tabular}{|c c c c|}
  \hline
  Crystal              & $\beta_\mathrm{mag}$ & $\beta_\mathrm{el}$ & $\beta_\mathrm{str}$   \\
  \hline
  Tm:YAG               & $6.9\cdot 10^{-17}$ & $2.4\cdot 10^{-18}$ & $2.3\cdot 10^{-12}$ \\
  Tm:YGG               & $1.3\cdot 10^{-18}$ & $2.1\cdot 10^{-18}$ & $1.9\cdot 10^{-12}$ \\
  Tm:LiNbO$_3$         & $1.1\cdot 10^{-15}$ & $3.0\cdot 10^{-14}$ & $1.5\cdot 10^{-12}$ \\
  Er:YSO               & $4.5\cdot 10^{-13}$ & $1.5\cdot 10^{-12}$ & $1.3\cdot 10^{-12}$ \\
  Er:LiNbO$_3$         & $1.7\cdot 10^{-13}$ & $5.8\cdot 10^{-14}$ & $1.5\cdot 10^{-12}$ \\
  Eu:YSO               & $6.7\cdot 10^{-20}$ & $7.4\cdot 10^{-13}$ & $1.3\cdot 10^{-12}$ \\
  Eu:Y$_2$O$_3$        & $6.7\cdot 10^{-20}$ & $4.9\cdot 10^{-13}$ & $1.3\cdot 10^{-12}$ \\
  EuCl$_3\cdot$6D$_2$O & $4.3\cdot 10^{-20}$ & $4.1\cdot 10^{-15}$ & $1.0\cdot 10^{-12}$ \\
  Pr:YSO               & $1.5\cdot 10^{-17}$ & $7.5\cdot 10^{-12}$ & $1.3\cdot 10^{-12}$ \\
  Pr:La$_2$(WO$_4$)$_3$ & $1.5\cdot 10^{-17}$ & $3.0\cdot 10^{-12}$ & $7.4 \cdot 10^{-13}$ \\
  \hline
\end{tabular}
\caption{Estimated values for the ISD coefficients $\beta_i$ in Hz\ cm$^3$.}
\label{tab:beta}
\end{table}

\section{Instantaneous Spectral Diffusion}
\label{sec:param}
In Table~\ref{tab:param} we list the physical parameters that are needed for the estimation of ISD strengths for an ensemble of 10 REIC. Note that some values had to be estimated using data measured in similar materials. The value for the piezospectroscopic sensitivity $\kappa$ being basically unknown for most REICs, we choose to take $\kappa=100$~Hz/Pa for all. This choice is supported by the observed similarity of the value of $\kappa$ in a broad variety of hosts, dopants and transitions~\cite{louchet2019piezospectroscopic}.

In Table~\ref{tab:beta} we present the different ISD coefficients $\beta_i$ calculated for the three possible ion-ion interactions (magnetic dipole-dipole interaction, electric dipole-dipole interaction, and strain-mediated interaction), using the parameters listed in Table~\ref{tab:param} and Eqs.~\ref{eq:Ael}, \ref{eq:Amag}  and \ref{eq:Astr}.

\section*{References}
\bibliographystyle{iopart-num}

\providecommand{\newblock}{}

\end{document}